\documentclass[preprint,review,3p]{elsarticle}

\usepackage{graphicx}
\usepackage{color}
\usepackage{soul}
\usepackage[linesnumbered, ruled]{algorithm2e}
\usepackage{subfigure}

\usepackage{threeparttable}
\usepackage{caption}
\usepackage{epstopdf}
\usepackage{amsthm}
\usepackage{amssymb, amsmath, amsfonts,  bm}
\DeclareMathOperator*{\argmin}{\arg\!\min}

\usepackage[utf8]{inputenc} 
\usepackage{titlesec}
\titleformat{\paragraph}
{\normalfont\normalsize\bfseries}{\theparagraph}{1em}{}
\titlespacing*{\paragraph}
{0pt}{3.25ex plus 1ex minus .2ex}{1.5ex plus .2ex}
\usepackage[final]{epsfig}
\usepackage{float}
\usepackage{array}
\usepackage{amsmath,amssymb,latexsym}
\usepackage{multirow}

\usepackage[usenames, dvipsnames]{xcolor}
\usepackage{pgfplots, tikz}
\usepackage{tkz-berge}
\usetikzlibrary{plotmarks, patterns, arrows, automata, petri, topaths}
\usetikzlibrary{shapes.geometric}

\begin{document}

\begin{frontmatter}

\author[label1,label2]{Virginie OLLIER \corref{cor1}}
\ead{virginie.ollier@satie.ens-cachan.fr}
\cortext[cor1]{Corresponding author}
\author[label3]{Mohammed Nabil EL KORSO}
\author[label4]{Andr\'{e} FERRARI}
\author[label2]{R\'{e}my BOYER}
\author[label1]{Pascal LARZABAL}
\address[label1]{ENS Paris-Saclay, SATIE UMR 8029, 61 avenue du Président Wilson, 94235 Cachan, France}
\address[label2]{Paris-Sud University, L2S UMR 8506, 3 rue Joliot-Curie, 91192 Gif-sur-Yvette, France}
\address[label3]{Paris-Nanterre University, LEME EA 4416, 50 rue de S\`evres, 92410 Ville d'Avray, France}
\address[label4]{Nice Sophia-Antipolis University, Lab. J.L. Lagrange, UMR 7293, Parc Valrose, 06108 Nice, France}
\tnotetext[t1]{This work was supported by MAGELLAN (ANR-14-CE23-0004-01), ON FIRE project (Jeunes Chercheurs GDR-ISIS) and ANR ASTRID project MARGARITA (ANR-17-ASTR-0015).}

\title{Robust distributed calibration of radio interferometers with direction dependent distortions}

\begin{abstract}

In radio astronomy, accurate calibration is of crucial importance for the new generation of radio interferometers. More specifically, because of the potential presence of outliers which affect the measured data, robustness needs to be ensured. On the other hand, calibration is improved by taking advantage of these new instruments and exploiting the known structure of parameters of interest across frequency. Therefore, we propose in this paper an iterative robust multi-frequency calibration algorithm based on a distributed and consensus optimization scheme which aims to estimate the complex gains of the receivers and the directional perturbations caused by the ionosphere. Numerical simulations reveal that the proposed distributed calibration technique outperforms the conventional non-robust algorithm and per-channel calibration.

\end{abstract}


\end{frontmatter}


\section{Introduction}
\label{Introduction}
In radio interferometry, the easiest approach to perform calibration is to consider one frequency bin at a time \cite{wijnholds2009multisource,wijnholds2010calibration}, with a single centralized agent processing the data, leading to suboptimality and computational limitations. However, considering the specific variation of parameters across frequency enables to add some modeling structure on the unknown parameters to estimate and therefore perform more accurately direction dependent calibration \cite{smirnov2011revisiting2}, notably for the new generation of radio interferometers where multiple frequency sub-bands are present \cite{deVos2009lofar}.
 Exploiting the frequency-dependent response has already been studied in the context of bandpass solutions \cite{trott2016spectral}, assuming a smooth response in frequency and also for direction dependent calibration with
smooth polynomials \cite{yatawatta2015distributed, martinjournalnew}. 
As direction dependent perturbation effects are particularly significant in observations with new instruments, we focus in this paper on the regime where the receiving elements of the array have a large field-of-view and possibly long baselines, resulting in a direction dependent calibration problem in which receptors see different parts of the ionosphere \cite{wijnholds2010calibration}. We call this the direction dependent distortion regime. \\
\indent On the one hand, multi-frequency calibration aims to take into account a whole frequency range, \textit{e.g.}, between 30 and 240 MHz for the nominal operating bandpass of the Low Frequency Array (LOFAR) \cite{van2013lofar, ferrari2014distributed}. A computationally efficient way to handle the multiple sub-frequency bands in radio astronomy is to apply distributed and consensus algorithms with a decentralized strategy. In this context, we apply the Alternating Direction Method of Multipliers (ADMM) \cite{boyd2011distributed,yatawatta2015distributed}, which is well-suited for large-scale problems as in radio interferometry.
This technique is based on decomposition and coordination tasks with a group of computational agents. Each of them has access to a part of the data and finds a solution to a local subproblem, in a restricted frequency interval. Communication with a fusion center enforces consensus among all agents, the goal being to solve a global constrained optimization problem. By collecting and storing data in a distributed way among different computational agents, the global operational cost in the network is substantially reduced. \\
\indent On the other hand, measurements are frequently affected by the presence of outliers due to interference or weak background sources. As a consequence, the noise can no longer be considered Gaussian as in \cite{yatawatta2015distributed}. In order to propose an alternative to \cite{yatawatta2014robust}, where the noise model is based on the Student's t distribution, in this work, we make use of
a compound-Gaussian model which includes heavy-tailed distributions \cite{ollila2012complex}. This is more general and reveals to be more robust \cite{ollierjournal}.  
Therefore, to take into account the presence of outliers and the variability of parameters across frequency, we propose here a Multi-frequency Robust Calibration Algorithm (MRCA) based on a compound-Gaussian distribution for modeling the noise and the consensus ADMM approach \cite{boyd2011distributed}. As in \cite{tasse2014nonlinear}, we aim at estimating physical parameters appearing in the Jones terms \cite{thompson2008interferometry,smirnov2011revisiting2}. \\ 
 \indent
In this paper, we use the following notation:  The trace and determinant operators  are, respectively, given by $\mathrm{tr}\left\{ \cdot \right\} $ and $| \cdot | $. The $B \times B$ identity matrix is referred by $\mathbf{I}_{B}$ and $||\cdot||_2$ denotes the $l_2$ norm. The symbol $\otimes$ represents the Kronecker product, $\mathrm{vec}(\cdot)$ stacks the columns of a matrix on top of one another, $\mathrm{diag}\{\cdot\}$ converts a vector into a diagonal matrix while $\mathrm{bdiag}\{\cdot\}$ is the block-diagonal operator. Finally, $\Re\left\{\cdot\right\}$ and $\Im\left\{\cdot\right\}$ are, respectively, the real and imaginary parts,  $[\cdot]_k$ refers to the \textit{k}-th entry of the considered vector, $\text{arg}\left\{ \cdot \right\} $ is the argument of a complex number and $j$ is the complex number whose square equals $-1$.

\section{Model setup}
\label{setup}

\subsection{Direction dependent distortion regime}
\label{regime}

An interferometer output consists of visibilities, \textit{i.e.}, correlations of signals measured by two array elements along the corresponding baseline vector. In the noise free case, the measurements are given, for each frequency $f \in \mathcal{F}=\{f_1,\hdots,f_F\}$, by
\begin{align}
\label{ME}
\mathbf{S}^{[f]}_{pq}(\boldsymbol{\theta}^{[f]}) = &
\sum_{i=1}^{D}\mathbf{J}^{[f]}_{i,p}(\boldsymbol{\theta}^{[f]})\mathbf{C}^{[f]}_{i}
\mathbf{J}_{i,q}^{[f]^H}(\boldsymbol{\theta}^{[f]})
\end{align}
where $(p,q)\in \{1, \ldots, M\}^2$ is a pair of receivers with $p<q$, $M$ denotes the total number of array elements and $D$ the number of calibrator sources, while $\mathbf{J}^{[f]}_{i,p}(\boldsymbol{\theta}^{[f]})$ refers to the so-called $2 \times 2$ Jones matrix \cite{thompson2008interferometry,smirnov2011revisiting2}, accounting for the perturbation effects along the path from the \textit{i}-th source to the \textit{p}-th receiver. We note $\boldsymbol{\theta}^{[f]}$ the unknown parameter vector of interest, whose elements are detailed further below. Finally, $\mathbf{C}^{[f]}_{i}$ is the known source brightness matrix of the \textit{i}-th calibrator source, describing its polarization state.
After vectorization of measurements (\ref{ME}) and considering the noise effect, we obtain the $4 \times 1$ data vector 
$\mathbf{x}^{[f]}_{pq}=
\mathrm{vec}\left(\mathbf{S}^{[f]}_{pq}(\boldsymbol{\theta}^{[f]})\right)+\mathbf{n}^{[f]}_{pq}$
where $\mathbf{n}^{[f]}_{pq}$ is the noise vector for baseline $(p,q)$ and 
$\mathrm{vec}\left(\mathbf{S}^{[f]}_{pq}(\boldsymbol{\theta}^{[f]})\right)=\sum_{i=1}^{D}\mathbf{s}^{[f]}_{i,pq}(\boldsymbol{\theta}^{[f]})$
in which $\mathbf{s}^{[f]}_{i,pq}(\boldsymbol{\theta}^{[f]})=\left(\mathbf{J}^{[f]^{\ast}}_{i,q}(\boldsymbol{\theta}^{[f]})\otimes
\mathbf{J}^{[f]}_{i,p}(\boldsymbol{\theta}^{[f]})\right)\mathbf{c}^{[f]}_{i}$ and $\mathbf{c}^{[f]}_{i}=\mathrm{vec}(\mathbf{C}^{[f]}_{i})$. Finally, the $4B \times 1$ full measurement vector at frequency $f$, with $B=\frac{M(M-1)}{2}$ the total number of baselines, reads
\begin{equation}
\label{express_x}
\mathbf{x}^{[f]}=\left[\mathbf{x}^{[f]^{T}}_{12},
\ldots,
\mathbf{x}^{[f]^{T}}_{(M-1)M}\right]^{T}
=\sum_{i=1}^{D}\left[\mathbf{s}^{[f]^T}_{i,12}(\boldsymbol{\theta}^{[f]}),
\ldots,
\mathbf{s}^{[f]^T}_{i,(M-1)M}(\boldsymbol{\theta}^{[f]})\right]^{T}+\left[\mathbf{n}^{[f]^T}_{12},
\ldots,
\mathbf{n}^{[f]^T}_{(M-1)M}\right]^{T}.
\end{equation}

In the direction dependent distortion regime, a particular decomposition of the Jones matrix is given by \cite{thompson2008interferometry,noordam2010meqtrees}
\begin{equation}
 \label{model_regime3}
\mathbf{J}^{[f]}_{i,p}(\boldsymbol{\theta}^{[f]})=\mathbf{G}^{[f]}_p(\mathbf{g}^{[f]}_p)\mathbf{H}^{[f]}_{i,p}\mathbf{Z}^{[f]}_{i,p}(\varphi^{[f]}_{i,p})\mathbf{F}^{[f]}_{i,p}(\vartheta^{[f]}_{i,p}).
\end{equation}
Specifically, the complex electronic gain matrix is represented by $\mathbf{G}^{[f]}_p(\mathbf{g}^{[f]}_p)=\mathrm{diag}\{\mathbf{g}^{[f]}_p\}$ while $\mathbf{H}^{[f]}_{i,p}$ is an assumed known matrix gathering the geometric delay and beam pattern \cite{thompson2008interferometry,smirnov2011revisiting2}.
In addition, propagation through the ionosphere induces two effects. The first one is a phase delay given by the matrix $\mathbf{Z}^{[f]}_{i,p}(\varphi^{[f]}_{i,p})$ \cite{smirnov2011revisiting2}, and written as 
  \begin{equation}
  \label{form_Z}
  \mathbf{Z}^{[f]}_{i,p}(\varphi^{[f]}_{i,p})=\exp\Big(j\varphi^{[f]}_{i,p}\Big)\mathbf{I}_2
  \end{equation}
 where $\varphi^{[f]}_{i,p} \propto \mathrm{TEC}_{i,p}/ f$ \cite{van2007ionospheric}, with $\mathrm{TEC}_{i,p}$ the Total Electron Content defined as the integrated electron density along line of sight $i$-$p$.
The second effect is the so-called Faraday rotation \cite{noordam2010meqtrees}  in (\ref{model_regime3}), given by 
\begin{equation}
\label{form_F}
\mathbf{F}^{[f]}_{i,p}(\vartheta^{[f]}_{i,p}) = \begin{bmatrix}
    \cos(\vartheta^{[f]}_{i,p}) & -\sin(\vartheta^{[f]}_{i,p}) \\
   \sin(\vartheta^{[f]}_{i,p
   }) & \cos(\vartheta^{[f]}_{i,p})
\end{bmatrix}
\end{equation}
where the unknown rotation angle $\vartheta^{[f]}_{i,p} \propto \mathrm{RM}_{i,p} / f^{2}$ \cite{smirnov2011revisiting2} and $\rm{RM}_{i,p}$ is the Rotation Measure which depends on the magnetic field and the electron density along the path $i$-$p$.  

Therefore, the $ (2 M D + 2 M) \times 1$ complex unknown parameter vector is given by $\boldsymbol{\theta}^{[f]}=[\boldsymbol{\epsilon}^{[f]^T},\mathbf{g}_1^{[f]^T},\hdots,\mathbf{g}_M^{[f]^T}]^T$
in which
$\boldsymbol{\epsilon}^{[f]}=[\vartheta^{[f]}_{1,1},\hdots, \vartheta^{[f]}_{D,M},\exp(j\varphi^{[f]}_{1,1}),\hdots, 
\exp(j\varphi^{[f]}_{D,M})]^T$ refers to the frequency dependent per-receiver and per-source ionospheric effects.

\subsection{Noise modeling as a compound-Gaussian distribution}
 
 The presence of outliers has multiple causes in radio astronomy, such as errors in the sky model due to weak sources in the background \cite{yatawatta2014robust} or man-made Radio Frequency Interference (RFI) \cite{raza2002spatial}, leading to statistics, with heavy-tailed distributions, different from the classical Gaussian case \cite{wijnholds2009multisource}. 
To ensure robustness in the proposed estimator, we adopt a two-scale compound-Gaussian noise modeling given for each baseline by
\begin{equation}
\label{sirp}
\mathbf{n}^{[f]}_{pq} = \sqrt{\tau^{[f]}_{pq}} \  \boldsymbol{\mu}^{[f]}_{pq},
\end{equation} 
where the power factor $\tau^{[f]}_{pq}$ is a positive real random variable and the $4 \times 1$ vector $\boldsymbol{\mu}^{[f]}_{pq}$ follows a zero-mean complex circular Gaussian distribution, \textit{i.e.}, $\boldsymbol{\mu}^{[f]}_{pq} \sim \mathcal{CN}(\mathbf{0},\boldsymbol{\Omega}^{[f]})$ \footnotemark.
 Therefore,
calibration amounts to estimate for each frequency $f$ the $ (2 M D + 2 M ) \times 1$ vector $\boldsymbol{\theta}^{[f]}$ describing Jones matrices, as well as $ B \times 1$ texture realizations $\boldsymbol{\tau}^{[f]}=[\tau^{[f]}_{12},\tau^{[f]}_{13},\ldots,\tau^{[f]}_{(M-1)M}]^{T}$ and the $4 \times 4$ speckle covariance matrix $\boldsymbol{\Omega}^{[f]}$ (which must satisfy, \textit{e.g.}, $\mathrm{tr}\left\{\boldsymbol{\Omega}^{[f]} \right\}=1$ to avoid the ambiguity with the power factor, the choice of this constraint being arbitrary \cite{zhang2016mimo}). In the following, we assume independence of $\mathbf{n}^{[f]}_{pq}$ between baselines and frequencies and no specific prior structure exists \textit{w.r.t.} $f$. Let us note that the algorithm can be adapted to perform an independent estimation of unknown parameters  for each time interval. In order to consider different time scales for the unknown effects, a similar approach to the one proposed in this work for the multi-frequency scenario can be adopted, using specific time variation models or assuming smoothness across time.

\footnotetext{It is possible to consider a baseline dependent covariance matrix $\boldsymbol{\Omega}^{[f]}_{pq}$. In this case, the proposed algorithm requires a few modifications  which are straightforward.
}

\section{Description of the proposed estimator}

In this section, we introduce the Relaxed Maximum Likelihood (RML) method. Then, the ADMM algorithm is proposed to estimate the frequency dependent parameters in a distributed way. In the iterative procedure, each subset of parameters is updated alternatively, while fixing the remaining parameters.

\subsection{Robust estimation of Jones matrices}

Robust calibration is based on the model (\ref{express_x}) and the compound-Gaussian noise model (\ref{sirp}). Estimations are performed iteratively with the ML method similar to that of \cite{zhang2016mimo}. Specifically, here, we choose not to specify the probability density function (pdf) of the texture parameters which are assumed unknown and deterministic, leading to the RML. By doing so, we ensure more flexibility and robustness to any prior mismatch \textit{w.r.t.} the unknown pdf of $\boldsymbol{\tau}^{[f]}$. 
Using independency between measurements, the log-likelihood function of the RML reads
\begin{equation}
\label{likelihood}
 \log f \left(\{\mathbf{x}^{[f]}\}_{f \in \mathcal{F}} \ | \ \{\boldsymbol{\theta}^{[f]}, \boldsymbol{\tau}^{[f]},\boldsymbol{\Omega}^{[f]}\}_{f \in \mathcal{F}}\right) 
=-\sum_{f \in \mathcal{F}}l^{[f]}(\boldsymbol{\theta}^{[f]})- \sum_{f \in\mathcal{F}} \sum_{pq} \log | \pi
\tau^{[f]}_{pq} \boldsymbol{\Omega}^{[f]} |
\end{equation}
in which $l^{[f]}(\boldsymbol{\theta}^{[f]})=\sum\limits_{pq}\frac{1}{\tau^{[f]}_{pq}}\mathbf{u}^{[f]^{H}}_{pq}(\boldsymbol{\theta}^{[f]})
\boldsymbol{\Omega}^{[f]^{-1}}\mathbf{u}^{[f]}_{pq}(\boldsymbol{\theta}^{[f]})$ with
$\mathbf{u}^{[f]}_{pq}(\boldsymbol{\theta}^{[f]}) = \mathbf{x}^{[f]}_{pq}- \sum\limits_{i=1}^{D}\mathbf{s}^{[f]}_{i,pq}(\boldsymbol{\theta}^{[f]})$. Let us recall that for a given frequency $f$, $\boldsymbol{\theta}^{[f]}$ is the $ (2 M D + 2 M ) \times 1$ vector of interest, $\boldsymbol{\tau}^{[f]}$ contains the $ B$ texture parameters and $\boldsymbol{\Omega}^{[f]}$ is the $4 \times 4$ speckle covariance matrix.
After some calculus, the optimization of the log-likelihood function \textit{w.r.t.} the noise parameters leads to the following texture estimate
\begin{equation}
\label{tauExp} \hat{\tau}^{[f]}_{pq} = \frac{1}{4} \mathbf{u}^{[f]^{H}}_{pq}(\boldsymbol{\theta}^{[f]})
\boldsymbol{\Omega}^{[f]^{-1}}\mathbf{u}^{[f]}_{pq}(\boldsymbol{\theta}^{[f]})
\end{equation}
while the speckle covariance matrix reads
\begin{equation}
\label{OmegaEstimBefore}
\hat{\boldsymbol{\Omega}}^{[f]}= \frac{4}{B}
\sum\limits_{pq} \frac{\mathbf{u}^{[f]}_{pq}(\boldsymbol{\theta}^{[f]})
\mathbf{u}^{[f]^H}_{pq}(\boldsymbol{\theta}^{[f]})}{\mathbf{u}^{[f]^H}_{pq}(\boldsymbol{\theta}^{[f]})\hat{\boldsymbol{\Omega}}^{[f]^{-1}}\mathbf{u}^{[f]}_{pq}(\boldsymbol{\theta}^{[f]})}
\end{equation}
followed by a normalization step, $\hat{\boldsymbol{\Omega}}^{[f]}  \leftarrow \frac{\hat{\boldsymbol{\Omega}}^{[f]}}{\mathrm{tr}\left\{\hat{\boldsymbol{\Omega}}^{[f]}\right\}}$, to avoid any ambiguity.
Consequently, the proposed algorithm consists in updating alternatively $\{\hat{\boldsymbol{\theta}}^{[f]}\}_{f \in \mathcal{F}}$, $\{\hat{\boldsymbol{\Omega}}^{[f]}\}_{f \in \mathcal{F}}$ and $\{\hat{\boldsymbol{\tau}}^{[f]}\}_{f \in \mathcal{F}}$ in a global iterative loop. Estimation of $\{\boldsymbol{\theta}^{[f]}\}_{f \in \mathcal{F}}$ can be achieved iteratively, for a fixed $\{\hat{\boldsymbol{\tau}}^{[f]}\}_{f \in \mathcal{F}}$ and $\{\hat{\boldsymbol{\Omega}}^{[f]}\}_{f \in \mathcal{F}}$, thanks to the ADMM procedure. More specifically, we need to maximize successively the cost funtion \textit{w.r.t.} the complex gain vector $\mathbf{g}^{[f]}=[\mathbf{g}_1^{[f]^T},\hdots,\mathbf{g}_M^{[f]^T}]^T$ and on the other hand, \textit{w.r.t.} the  direction dependent ionospheric effects $\boldsymbol{\epsilon}^{[f]}$.

\subsection{Distributed estimation of the frequency dependent parameters}

To estimate the parameter vector $\{\boldsymbol{\theta}^{[f]}\}_{f \in \mathcal{F}}$, with $\{\hat{\boldsymbol{\tau}}^{[f]}\}_{f \in \mathcal{F}}$ and $\{\hat{\boldsymbol{\Omega}}^{[f]}\}_{f \in \mathcal{F}}$ fixed,  we use the ADMM procedure \cite{boyd2011distributed}, which is adapted for distributed optimization problems. In such case, a set of computational agents is considered. Each of them achieves calibration locally for a specific frequency $f$ and transmits the estimates to the fusion center where consensus among all agents is enforced thanks to known constraints. 
Closed-form expressions are not straightforward to obtain, in order to overcome this difficulty, we rewrite the ionospheric phase delay in (\ref{form_Z}), for $\ i \in \{1,\ldots,D\}$ and  $\ p \in \{1,\ldots,M\}$, as 
$\exp\Big(j\varphi^{[f]}_{i,p}\Big)=\mathbf{b}^{[f]^T}\mathbf{z}_{i,p}$
and the Faraday rotation angle in (\ref{form_F}) as 
$\vartheta^{[f]}_{i,p}=\frac{1}{f^2}\overline{z}_{i,p}$
\textit{s.t.} $\mathbf{z}_i=[\overline{z}_{i,1},\hdots,\overline{z}_{i,M},\mathbf{z}^T_{i,1},\hdots,\mathbf{z}^T_{i,M}]^T$ is the vector of hidden variables for the \textit{i}-th source direction, which are frequency independent and need to be estimated. 
The $N \times 1$ vector $\mathbf{b}^{[f]}$ is defined \textit{s.t.} $[\mathbf{b}^{[f]}]_{k}=\left(\frac{1}{f}\right)^{k-1}$ where $N-1$ is the order of the truncated approximation power series of $\exp\Big(j\varphi^{[f]}_{i,p}\Big)$.
As regards the gains, we enforce smoothness with polynomials thanks to
\begin{equation}
\mathbf{g}^{[f]}_p=\tilde{\mathbf{B}}^{[f]} \tilde{\mathbf{z}}_p
\end{equation}
where $\tilde{\mathbf{B}}^{[f]}=\tilde{\mathbf{b}}^{[f]^T} \otimes \mathbf{I}_2$, in which the $ \tilde{N} \times 1$ vector $\tilde{\mathbf{b}}^{[f]}$ is defined \textit{s.t.} $[\tilde{\mathbf{b}}^{[f]}]_k=\left(\frac{f-f_0}{f_0}\right)^{k-1}$ for a given reference frequency $f_0$  \cite{yatawatta2015distributed}. 
 Finally, the $2 \tilde{N} \times 1$ vector $\tilde{\mathbf{z}}_p$ contains the corresponding hidden variables for the \textit{p}-th receiving element.
Consequently, multi-frequency calibration is formulated as 
  \begin{equation}
   \{\hat{\boldsymbol{\theta}}^{[f]}\}_{f \in \mathcal{F}}, \hat{\mathbf{z}}  = \argmin_{\{\boldsymbol{\theta}^{[f]}\}_{f \in \mathcal{F}},\mathbf{z}} \sum_{f \in \mathcal{F}} l^{[f]}(\boldsymbol{\theta}^{[f]}) \ \ 
   \nonumber \text{s.t.} \ \  \boldsymbol{\theta}^{[f]}= \mathbf{B}^{[f]}\mathbf{z}
   \end{equation}
 where $\mathbf{z}=[\overline{z}_{1,1},\hdots,\overline{z}_{D,M},\mathbf{z}^T_{1,1}\hdots,\mathbf{z}^T_{D,M},\tilde{\mathbf{z}}^T_1,\hdots,\tilde{\mathbf{z}}^T_M]^T$  is the  augmented vector of hidden variables and the global frequency basis matrix is given by 
$\mathbf{B}^{[f]}= \mathrm{bdiag}\{\frac{1}{f^2}\mathbf{I}_{D M},\mathbf{I}_{D M} \otimes \mathbf{b}^{[f]^T},\mathbf{I}_M \otimes  \tilde{\mathbf{B}}^{[f]}\}.$
Thus, consensus optimization is associated with the following augmented Lagrangian problem \cite{boyd2011distributed}
   $ L\left( \{\boldsymbol{\theta}^{[f]}\}_{f \in \mathcal{F}},\mathbf{z},\{\mathbf{y}^{[f]}\}_{f \in \mathcal{F}}\right) 
   = \sum_{f \in \mathcal{F}} L^{[f]}\left(\boldsymbol{\theta}^{[f]},\mathbf{z},\mathbf{y}^{[f]}\right)$,
where $L^{[f]}\left(\boldsymbol{\theta}^{[f]},\mathbf{z},\mathbf{y}^{[f]}\right)= l^{[f]}\left(\boldsymbol{\theta}^{[f]} \right)+  h^{[f]}\left(\boldsymbol{\theta}^{[f]},\mathbf{z},\mathbf{y}^{[f]}\right)$, in which using \cite{LiADMMcomplex, jiang2014alternating}, it can be proven that
\begin{equation}
h^{[f]}\left(\boldsymbol{\theta}^{[f]},\mathbf{z},\mathbf{y}^{[f]}\right)= 
2\Re\left\{\mathbf{y}^{[f]^H}\left(\boldsymbol{\theta}^{[f]}-\mathbf{B}^{[f]}\mathbf{z} \right)\right\}+\rho||\boldsymbol{\theta}^{[f]}-\mathbf{B}^{[f]}\mathbf{z}||^2_2
\end{equation}
and $\mathbf{y}^{[f]} =[\overline{y}_{1,1}^{[f]},\hdots,\overline{y}_{D,M}^{[f]},y_{1,1}^{[f]},\hdots,y_{D,M}^{[f]},\tilde{\mathbf{y}}^{[f]^T}_1,\hdots,\tilde{\mathbf{y}}^{[f]^T}_M]^T$ denotes the associated Lagrange multipliers for a given $f$ while $\rho$ is a regularization factor. Due to separability in frequency, calibration can be carried out independently by each agent.
The three steps of the ADMM algorithm are the following for the \textit{t}-th iteration 
 \begin{equation}
 \label{step1}
   \bullet \ \  \left(\hat{\boldsymbol{\theta}}^{[f]}\right)^{t+1} =  \argmin_{\boldsymbol{\theta}^{[f]}}L^{[f]}\left(\boldsymbol{\theta}^{[f]},\left(\hat{\mathbf{z}}\right)^{t},\left(\hat{\mathbf{y}}^{[f]}\right)^{t}\right)  
    \ \text{performed locally by each agent}
      \end{equation}
      \begin{equation}
       \label{step2}
     \bullet \ \  \left(\hat{\mathbf{z}}\right)^{t+1}  =\argmin_{\mathbf{z}} \sum_{f \in \mathcal{F}} h^{[f]}\left(\left(\hat{\boldsymbol{\theta}}^{[f]}\right)^{t+1},\mathbf{z},\left(\hat{\mathbf{y}}^{[f]}\right)^{t}\right)     
    \ \text{performed globally at the fusion center}
      \end{equation}
     \begin{equation}
      \label{step3}
    \bullet \ \  \left(\hat{\mathbf{y}}^{[f]}\right)^{t+1} =\left(\hat{\mathbf{y}}^{[f]}\right)^{t}+ \rho\left(\left(\hat{\boldsymbol{\theta}}^{[f]}\right)^{t+1} -  \mathbf{B}^{[f]}\left(\hat{\mathbf{z}}\right)^{t+1}\right)     
    \ \text{performed locally by each agent}
      \end{equation}
      Solving (\ref{step2}) directly leads to the following closed-form expression
\begin{equation}
\label{c_p}
\hat{\mathbf{z}} = \left(\rho\sum_{f \in \mathcal{F}} \Big(\mathbf{B}^{[f]^H}\mathbf{B}^{[f]} \Big)\right)^{-1} \left(\sum_{f \in \mathcal{F}} \mathbf{B}^{[f]^T}\Big( \mathbf{y}^{[f]}+ \rho\boldsymbol{\theta}^{[f]} \Big)\right)
\end{equation}       
which is then broadcasted to all agents as a common global variable.
Minimization (\ref{step1}) is performed thanks to an iterative approach. Specifically, we minimize \textit{w.r.t.} $\boldsymbol{\epsilon}^{[f]}$ on the one hand, which is done by minimizing successively \textit{w.r.t.} each
$\vartheta^{[f]}_{i,p}$ and $\varphi^{[f]}_{i,p}$ for $i \in \{1,\hdots,D\}$ and $p \in \{1,\hdots,M\}$, and \textit{w.r.t.} $\mathbf{g}^{[f]}$ on the other hand.
Estimation of $\vartheta^{[f]}_{i,p}$ is computed thanks to a 1D Newton or gradient descent-type algorithm \cite{nocedal2006numerical}, possibly in parallel for $i \in\{1,\hdots,D\}$ and $p \in\{1,\hdots,M\}$, and requires to solve 
\begin{equation}
\label{cas_faraday}
\hat{\vartheta}^{[f]}_{i,p}=\argmin_{\vartheta^{[f]}_{i,p} }
L^{[f]}\left(\boldsymbol{\theta}^{[f]},\mathbf{z},\mathbf{y}^{[f]}\right).
\end{equation}    
Regarding the ionospheric phase delays,
 minimization (\ref{step1}) for a given $\hat{\vartheta}_{i,p}^{[f]}$ for $ i \in \{1,\hdots,D\}$ and $ p \in \{1,\hdots,M\}$ leads to 
\begin{equation}
\label{express_phi}
\hat{\varphi}^{[f]}_{i,p}=\frac{1}{2} \text{arg}\left\{-\frac{\alpha^{[f]}_{i,p}}{\beta^{[f]}_{i,p}}\right\},
\end{equation}
 for which notations are specified in Appendix A.
For sake of clarity, in the following, dependence \textit{w.r.t.} parameters of interest is dropped and we recall that $\mathbf{g}^{[f]}_p=[[\mathbf{g}^{[f]}_p]_1,[\mathbf{g}^{[f]}_p]_2]^T$. The first-order derivative of the cost function (\ref{likelihood}) \textit{w.r.t.} $[\mathbf{g}^{[f]}_p]_1$ leads, after some calculus, to
\begin{equation}
\label{cost_gain}
\frac{\partial l^{[f]} (\mathbf{g}^{[f]})}{\partial [\mathbf{g}^{[f]}_p]_1}= \sum_{\substack {q=1 \\ q>p}}^M\frac{1}{\tau^{[f]}_{pq}}  \lambda^{[f]}_{pq}+
 \sum_{\substack {q=1 \\ q < p}}^M\frac{1}{\tau^{[f]}_{qp}} \tilde{\lambda}^{[f]}_{qp}
\end{equation}
in which the expressions of $\lambda^{[f]}_{pq}$ and $\tilde{\lambda}^{[f]}_{qp}$ are given in Appendix B.
After tedious calculations and rearrangement, we obtain the following solution	
\begin{equation}
\label{last_cost_gain}
[\mathbf{g}^{[f]}_p]_1=\frac{a^{[f]}_p}{b^{[f]}_p}
\end{equation}
where $a^{[f]}_p=-t^{^{[f]^\ast}}_p - [\mathbf{g}^{[f]}_p]_2\left(\sum\limits_{k=1}^{2S}[\mathbf{w}^{[f]}_p]_{2k}[\boldsymbol{\omega}^{^{[f]^\ast}}_p]_{2k} +\sum\limits_{k=1}^{2V}[\boldsymbol{\varsigma}^{[f]}_p]_{2k} [\boldsymbol{\varrho}^{[f]^{\ast}}_p ]_{2k}\right)-[\tilde{\mathbf{y}}_{p}^{[f]}]_1 +\rho \tilde{\mathbf{z}}^{T}_{p}[\tilde{\mathbf{B}}^{[f]^{T}}]_{:,1}$ and $b^{[f]}_p=\sum\limits_{k=0}^{2S -1}[\mathbf{w}^{[f]}_p]_{2k+1}[\boldsymbol{\omega}^{[f]^{\ast}}_p]_{2k+1}+\sum\limits_{k=0}^{2V -1}[\boldsymbol{\varsigma}^{[f]}_p]_{2k+1} [\boldsymbol{\varrho}^{[f]^{\ast}}_p ]_{2k+1}+\rho$, for which detailed notations are exposed in Appendix B. Estimation of $[\mathbf{g}^{[f]}_p]_2$ is performed in a similar way, leading to
\begin{equation}
\label{last_cost_gain2}
[\mathbf{g}^{[f]}_p]_2=\frac{\tilde{a}^{[f]}_p}{\tilde{b}^{[f]}_p}
\end{equation}
where $\tilde{a}^{[f]}_p=-t^{[f]^{\ast}}_p-[\mathbf{g}^{[f]}_p]_1\left(\sum\limits_{k=0}^{2S -1}[\mathbf{w}^{[f]}_p]_{2k+1}[\boldsymbol{\omega}^{[f]^{\ast}}_p]_{2k+1}+\sum\limits_{k=0}^{2V -1}[\boldsymbol{\varsigma}^{[f]}_p]_{2k+1} [\boldsymbol{\varrho}^{[f]^{\ast}}_p ]_{2k+1}\right)-[\tilde{\mathbf{y}}_{p}^{[f]}]_2+\rho \tilde{\mathbf{z}}^{T}_{p}[\tilde{\mathbf{B}}^{[f]^{T}}]_{:,2}$ and $\tilde{b}^{[f]}_p=\sum\limits_{k=1}^{2S}[\mathbf{w}^{[f]}_p]_{2k}[\boldsymbol{\omega}^{[f]^{\ast}}_p]_{2k}+\sum\limits_{k=1}^{2V}[\boldsymbol{\varsigma}^{[f]}_p]_{2k} [\boldsymbol{\varrho}^{[f]^{\ast}}_p ]_{2k}+\rho$. We introduce the \textit{Matlab} notation $[\cdot]_{:,k}$ to refer to the \textit{k}-th column of the corresponding matrix.
Finally, the global scheme of the proposed MRCA is described in Table 1 while the operation flow and signaling exchange between fusion center and local agents is shown in Figure 1.

\section{Numerical simulations}

In this part, we compare the performance of the proposed MRCA with a non-robust algorithm based on a Gaussian noise assumption and performed with an alternating least squares (ALS) method. This method consists in updating sequentially the parameters of interest in $\boldsymbol{\theta}^{[f]}$ for $f \in \mathcal{F}$, by minimizing a least squares cost function similar to that of \cite{wijnholds2009multisource,yatawatta2015distributed}.
We also consider a per-channel calibration which means that calibration is performed for each frequency separately without exploiting information across frequency while multi-frequency algorithm refers to distributed calibration obtained with the ADMM procedure. 
 We generate each electronic gain entry $[\mathbf{g}^{[f]}_p]_k$ as a complex circular Gaussian random variable with mean one and variance $\frac{1}{4}$, while ionospheric phase delay and Faraday rotation are randomly generated, as function of the $\rm{TEC}$ drawn from a uniform distribution $\mathcal{U}(1 \times 10^{17},5 \times 10^{17})$ and expressed in m$^{-2}$.
 We select $N=\tilde{N}=6$  and the extended Lagrangian parameter is set at $\rho=10$.  We consider $M=8$ receiving elements, with hundreds of meters for the longest baseline, $D=2$ bright signal sources and $D'=4$ weak background sources.
 Let us mention that we are considering here notional examples for the purpose of algorithm validation.
In Figure 2, we plot the Mean Square Error (MSE)  of the real part of one given parameter of vector $\mathbf{g}^{[f_1]}$ as a function of the Signal-to-Interference-plus-Noise Ratio (SINR), for different number of frequencies $F$ which are selected in the range $75-125$ MHz and $f_0=100$ MHz. The SINR is defined as the ratio of the normalized power of $D$ calibrator sources over the sum of normalized power of $D'$ background sources and a noise factor.  For each SINR, we perform $100$ Monte-Carlo runs. 
 A few iterations, less than $5$ for each loop in Table 1, are sufficient to attain stability in convergence, likewise for the so-called primal and dual residuals (approximately $5$ iterations),
  whose convergence mostly depends on parameters $\rho$, $N$, $\tilde{N}$ and initializations \cite{yatawatta2015distributed,boyd2011distributed}.
 Initial values for $\hat{\mathbf{y}}^{[f]}$ are chosen as a vector full of ones.
 Better estimation is achieved with the proposed algorithm and the behavior remains the same for other parameters of interest.
 
 In the following Figures, the MeqTrees software \cite{noordam2010meqtrees} is used to generate the observations and ionospheric effects need to be corrected in the presence of Gaussian background noise. We consider $M=7$ receptors, $D=2$ and $D'=16$, taken from the SUMSS survey with a spectral index of $0.7$. An image area of $3.5$ by $3.5$ degrees is shown in Figure 3, with and without the calibrator source, before adding any perturbation.  The integration time per visibility point is fixed at $60$s, the total synthesis time at $12$ hours and the SINR is approximately equal to 4 dB. After calibration, performed with MRCA and different classical methods, and subtraction of the bright calibrator from the data, we show the dirty image, \textit{i.e.}, the corrected residual in Figures 4 and 5, computed using \textit{lwimager}. Corresponding peak flux of the recovered weak sources in Figure 5 are shown in Table 2.
As expected, we notice better estimation performance and flux recovering when the proposed robust multi-frequency dependent calibration is performed in regards to the per-channel and non-robust cases, \textit{i.e.}, lower error and more robustness are achieved.

\section{Conclusion}

In this work, a multi-frequency robust iterative calibration algorithm based on compound-Gaussian modeling and the ADMM approach was described for the so-called direction dependent distortion regime.  Numerical simulations highlight the better estimation performance of the proposed technique \textit{w.r.t.} non-robust and/or per-channel cases, and its ability to accurately solve for per-direction and per-antenna $TEC$ and $RM$ physical quantities.

\appendix

\section{}

We specify here the notations which lead to (\ref{express_phi}). To this end, we need to consider $\frac{\partial l^{[f]}(\boldsymbol{\theta}^{[f]})}{\partial \varphi^{[f]}_{i,p}}$ and $\frac{\partial h^{[f]}(\boldsymbol{\theta}^{[f]})}{\partial \varphi^{[f]}_{i,p}}$. Thus, minimization (\ref{step1}) \textit{w.r.t.} phase delay $\varphi^{[f]}_{i,p}$ leads to
\begin{equation}
\label{before_final}
\beta^{[f]}_{i,p}\exp\Big(j\varphi^{[f]}_{i,p}\Big)+\alpha^{[f]}_{i,p}\exp\Big(-j\varphi^{[f]}_{i,p}\Big)=0
\end{equation}
in which
$\beta^{[f]}_{i,p}=\sum_{\substack {q=1 \\ q > p}}^M\frac{j}{\tau^{[f]}_{pq}}
  \exp\Big(-j\varphi^{[f]}_{i,q}\Big) \delta^{[f]}_{i,pq}+
\sum_{\substack {q=1 \\ q < p}}^M\frac{j}{\tau^{[f]}_{qp}}
\exp\Big(-j\varphi^{[f]}_{i,q}\Big)\delta^{[f]^H}_{i,qp} +j y_{i,p}^{[f]^{\ast}}-j\rho\mathbf{z}_{i,p}^H\mathbf{b}^{[f]^{\ast}}$
 and  
$ \alpha^{[f]}_{i,p}=\sum_{\substack {q=1 \\ q > p}}^M\frac{-j}{\tau^{[f]}_{pq}}
 \exp\Big(j\varphi^{[f]}_{i,q}\Big) \delta^{[f]^H}_{i,pq}-
\sum_{\substack {q=1 \\ q < p}}^M\frac{j}{\tau^{[f]}_{qp}}
\exp\Big(j\varphi^{[f]}_{i,q}\Big)\delta^{[f]}_{i,qp} 
-j y^{[f]}_{i,p}+j\rho\mathbf{b}^{[f]^T}\mathbf{z}_{i,p}$. 
We also introduce $\delta^{[f]}_{i,pq}=\left(-\mathbf{x}^{[f]^H}_{pq}+\sum\limits_{\substack {k=1\\ k\neq i}}^D\mathbf{s}^{[f]^H}_{k,pq}\right)
\boldsymbol{\Omega}^{[f]^{-1}} \mathbf{d}^{[f]}_{i,pq} $ and
$\mathbf{d}^{[f]}_{i,pq}=\left(\mathbf{G}^{[f]^{\ast}}_q(\mathbf{g}^{[f]}_q) \mathbf{H}^{[f]^{\ast}}_{i,q} \mathbf{F}^{[f]}_{i,q}(\vartheta^{[f]}_{i,q})\right) \otimes \left(\mathbf{G}^{[f]}_p(\mathbf{g}^{[f]}_p)
\otimes \mathbf{H}^{[f]}_{i,p}
 \mathbf{F}^{[f]}_{i,p}(\vartheta^{[f]}_{i,p})\right)\mathbf{c}^{[f]}_{i}$.
Finally, we directly deduce (\ref{express_phi}) from (\ref{before_final}).

\section{}
We describe here the notations introduced in the estimation of frequency dependent 
electronic gains. The expressions in (\ref{cost_gain}) are given by
\begin{equation}
\lambda^{[f]}_{pq}=t^{[f]}_{pq}+\mathbf{w}^{[f]^H}_{pq}
\left(\mathbf{I}_2  \otimes \mathbf{G}^{[f]^{\ast}}_{p}\right)
\boldsymbol{\varpi}^{[f]}_{pq}
\end{equation}
in which $t^{[f]}_{pq}=-\mathbf{x}^{[f]^H}_{pq}\mathbf{m}^{[f]}_{pq}$ with $\mathbf{m}^{[f]}_{pq}= \boldsymbol{\Omega}^{[f]^{-1}}
\left(\mathbf{G}^{[f]^{\ast}}_{q} \otimes 
\mathbf{E}_1
 \right)
\mathbf{w}^{[f]}_{pq}$ and $\mathbf{E}_1=\begin{bmatrix} 
1 & 0 \\
0 & 0
\end{bmatrix}$. Besides, $\mathbf{w}^{[f]}_{pq}=\sum\limits_{i=1}^D  \left(\mathbf{H}^{[f]^{\ast}}_{i,q} \mathbf{Z}^{[f]^{\ast}}_{i,q}(\varphi_{i,q}) \mathbf{F}^{[f]}_{i,q}(\vartheta^{[f]}_{i,q}) \right)    \otimes \left(\mathbf{H}^{[f]}_{i,p}
 \mathbf{Z}^{[f]}_{i,p}(\varphi_{i,p}) \\
\mathbf{F}^{[f]}_{i,p}(\vartheta^{[f]}_{i,p})\right)
\mathbf{c}^{[f]}_{i}$ and $
\boldsymbol{\varpi}^{[f]}_{pq}=
\left(\mathbf{G}^{[f]}_{q} \otimes \mathbf{I}_2 \right)
\mathbf{m}^{[f]}_{pq}$. 
Similarly, we have
\begin{equation}
\tilde{\lambda}^{[f]}_{qp}=\tilde{t}^{[f]}_{qp}+\tilde{\mathbf{w}}^{[f]^H}_{qp}
\left(\mathbf{G}^{[f]^{\ast}}_{p}\otimes \mathbf{I}_2 \right)
\tilde{\boldsymbol{\varpi}}^{[f]}_{qp}
\end{equation}
in which $\tilde{t}^{[f]}_{qp}=-\tilde{\mathbf{w}}^{[f]^H}_{qp}\mathbf{x}^{[f]}_{qp}$
with
$\tilde{\mathbf{w}}^{[f]}_{qp}=
\boldsymbol{\Omega}^{[f]^{-1}}
\left(
\mathbf{E}_1
\otimes  \mathbf{G}^{[f]}_{q} 
 \right)
\mathbf{w}^{[f]}_{qp}$
 and  $\tilde{\boldsymbol{\varpi}}^{[f]}_{qp}=
\left(\mathbf{I}_2  \otimes \mathbf{G}^{[f]}_{q}\right)
\mathbf{w}^{[f]}_{qp}$. 
The derivative in (\ref{cost_gain}) can be written more compactly as
\begin{equation}
\frac{\partial l^{[f]}(\mathbf{g}^{[f]})}{\partial [\mathbf{g}^{[f]}_p]_1}=t^{[f]}_p+\mathbf{w}^{[f]^H}_p\left(\mathbf{I}_S\otimes\mathbf{I}_2\otimes\mathbf{G}^{[f]^{\ast}}_{p}\right)\boldsymbol{\omega}^{[f]}_p+\tilde{\mathbf{w}}^{[f]^H}_p\left(\mathbf{I}_V\otimes\mathbf{G}^{[f]^{\ast}}_{p}\otimes\mathbf{I}_2\right)\tilde{\boldsymbol{\omega}}^{[f]}_p
\end{equation}
in which 
$\mathbf{w}_p^{[f]}=[\frac{1}{\tau^{[f]}_{p(p+1)}}\mathbf{w}^{[f]^T}_{p(p+1)}, \hdots,\frac{1}{\tau^{[f]}_{pM}}\mathbf{w}^{[f]^T}_{pM}]^T$,
$\tilde{\mathbf{w}}^{[f]}_{p}=[\frac{1}{\tau^{[f]}_{1p}}\tilde{\mathbf{w}}^{[f]^T}_{1p},\hdots,\frac{1}{\tau^{[f]}_{(p-1)p}}\tilde{\mathbf{w}}^{[f]^T}_{(p-1)p}]^T$
and 
$t^{[f]}_p= \sum\limits_{\substack {q=1 \\ q > p}}^M\frac{1}{\tau^{[f]}_{pq}}t^{[f]}_{pq}+\sum\limits_{\substack {q=1 \\ q < p}}^M\frac{1}{\tau^{[f]}_{qp}}\tilde{t}^{[f]}_{qp}$.
Similarly, we introduce $\boldsymbol{\omega}^{[f]}_p=[\boldsymbol{\varpi}^{[f]^T}_{p(p+1)},\hdots,\boldsymbol{\varpi}^{[f]^T}_{pM}]^T$ and  $\tilde{\boldsymbol{\omega}}^{[f]}_p=[\tilde{\boldsymbol{\varpi}}^{[f]^T}_{1p},\hdots,\tilde{\boldsymbol{\varpi}}^{[f]^T}_{(p-1)p}]^T$. 
 If we are considering the \textit{p}-th array element, we note $S=(M-p)$ and $V=(p-1)$.
Let us define the permutation matrix $\mathbf{P} = [\mathbf{e}_1,\mathbf{e}_3,\mathbf{e}_2,\mathbf{e}_4]$
where the $4 \times 1$ vector  $\mathbf{e}_k$ has zeros except at the \textit{k}-th position which is equal to unity.
With $\boldsymbol{\varsigma}^{[f]}_p=\left( \mathbf{I}_V \otimes \mathbf{P}^T\right)\tilde{\mathbf{w}}^{[f]}_p$ and $\boldsymbol{\varrho}^{[f]}_p=\left(\mathbf{I}_V \otimes \mathbf{P} \right)\tilde{\boldsymbol{\omega}}^{[f]}_p$, we obtain the following 1D linear equation 
\begin{align}
\label{ref_sur_g}
\nonumber
\frac{\partial l^{[f]}(\mathbf{g}^{[f]})}{\partial [\mathbf{g}^{[f]}_p]_1} = & t^{[f]}_p+[\mathbf{g}^{[f]^{\ast}}_p]_1\sum\limits_{k=0}^{2S -1}[\mathbf{w}^{[f]^{\ast}}_p]_{2k+1}[\boldsymbol{\omega}^{[f]}_p]_{2k+1}+[\mathbf{g}^{[f]^{\ast}}_p]_1\sum\limits_{k=0}^{2V -1}[\boldsymbol{\varsigma}^{[f]^{\ast}}_p]_{2k+1} [\boldsymbol{\varrho}^{[f]}_p ]_{2k+1} \\ &
+[\mathbf{g}^{[f]^{\ast}}_p]_2\sum\limits_{k=1}^{2S}[\mathbf{w}^{[f]^{\ast}}_p]_{2k}[\boldsymbol{\omega}^{[f]}_p]_{2k}+[\mathbf{g}^{[f]^{\ast}}_p]_2\sum\limits_{k=1}^{2V}[\boldsymbol{\varsigma}^{[f]^{\ast}}_p]_{2k} [\boldsymbol{\varrho}^{[f]}_p]_{2k}.
\end{align}
Considering (\ref{ref_sur_g}) and $\frac{\partial  h^{[f]}(\mathbf{g}^{[f]})}{\partial [\mathbf{g}^{[f]}_p]_1}$ leads to (\ref{last_cost_gain}).  
Estimation of $[\mathbf{g}^{[f]}_{p}]_{2}$ is similar, except that \\ $\mathbf{m}^{[f]}_{pq}= \boldsymbol{\Omega}^{[f]^{-1}}
\left(\mathbf{G}^{[f]^{\ast}}_{q} \otimes 
\mathbf{E}_2
\right)
\mathbf{w}^{[f]}_{pq}$ 
and 
$\tilde{\mathbf{w}}^{[f]}_{qp}=
 \boldsymbol{\Omega}^{[f]^{-1}}
  \left(
 \mathbf{E}_2
 \otimes \mathbf{G}^{[f]}_{q} 
 \right)
 \mathbf{w}^{[f]}_{qp}$ where $\mathbf{E}_2=\begin{bmatrix} 
 0 & 0 \\
 0 & 1
\end{bmatrix} $.

\bibliographystyle{elsarticle-num}
\bibliography{biblio}

\newpage
\begin{algorithm}
\SetAlgorithmName{Table 1}{}{} \caption{MRCA}
\SetKwInOut{input}{input} \SetKwInOut{output}{output}
\SetKwInOut{initialize}{initialize}
\initialize{$\{\hat{ \boldsymbol{\theta}}^{[f]} \leftarrow \boldsymbol{\theta}^{[f]}_{\mathrm{init}}\}_{f \in \mathcal{F}}$, $\hat{\mathbf{z}}$ $\leftarrow$ $\mathbf{z}_{\mathrm{init}}$, $\{\hat{\mathbf{y}}^{[f]}\ \leftarrow \mathbf{y}^{[f]}_{\mathrm{init}}\}_{f\in \mathcal{F}}$, $\{\hat{ \boldsymbol{\Omega}}^{[f]} \leftarrow \boldsymbol{\Omega}^{[f]}_{\mathrm{init}}\}_{f \in \mathcal{F}}$,$\{\hat{\tau}^{[f]}_{pq} \leftarrow \tau^{[f]}_{pq_{\mathrm{init}}}\}_{f \in \mathcal{F},  \ p<q, \ (p,q) \in \{1,\hdots,M\}^2}$}
\While
{stop criterion unreached}
{
\While 
{stop criterion unreached}
{
\While   
{ stop criterion unreached}{
\ShowLn Obtain $\{\hat{\vartheta}_{i,p}^{[f]}\}_{i=1,\hdots,D, \  p=1,\hdots,M} $ locally  from (\ref{cas_faraday})\\
\ShowLn Obtain $\{\hat{\varphi}_{i,p}^{[f]}\}_{i=1,\hdots,D, \  p=1,\hdots,M}$ locally from (\ref{express_phi})\\
\ShowLn Obtain $\hat{\mathbf{g}}^{[f]}$ locally from (\ref{last_cost_gain}) and (\ref{last_cost_gain2}) \\
}
\ShowLn Obtain $\hat{\mathbf{z}}$ globally from (\ref{c_p})\\
\ShowLn Obtain $\hat{\mathbf{y}}^{[f]}$ locally from (\ref{step3})\\
}
\ShowLn Obtain $\{\hat{\boldsymbol{\Omega}}^{[f]}\}_{f\in \mathcal{F}}$ from (\ref{OmegaEstimBefore})\\
\ShowLn Obtain $\{\hat{\tau}^{[f]}_{pq}\}_{f\in \mathcal{F}, \  p<q, \ (p,q) \ \in \ \{1,\hdots,M\}^2}$ from (\ref{tauExp})\\
}
\end{algorithm}

\vspace{1cm}

\begin{figure}[h!]

\centering

\tikzstyle{block} = [rectangle, draw,text width=7em, text centered, rounded corners, minimum height=5.5cm]
	\tikzstyle{doubleLine} = [draw, -latex,double,shorten <=1pt,shorten >=1pt]
	\tikzstyle{line} = [thick,draw, -latex',shorten <=1pt,shorten >=1pt]
	\tikzstyle{line2} = [thick,draw, latex'-,shorten <=1pt,shorten >=1pt]
	\tikzstyle{cloud} = [node distance=1cm]

	\begin{tikzpicture}[node distance = 3cm, auto]
	\node [cloud,transform canvas={yshift=0cm}] (input) {input: $\left\{\mathbf{x}^{[f]},\hat{\boldsymbol{\theta}}^{[f]},\hat{\boldsymbol{\Omega}}^{[f]}\right\}_{f \in \mathcal{F}}$, $\{\hat{\tau}^{[f]}_{pq}\}_{f \in \mathcal{F},  \ p<q, \ (p,q) \in \{1,\hdots,M\}^2}$};    
	\node [block, below of=input,node distance = 3.5cm] (localAgent) {\textit{k}-th local processor for frequency $f_k$};
	\node [block, right of=localAgent,node distance=10.5cm] (globalAgent) {fusion center};
	\node [cloud, below of=localAgent,node distance = 3.5cm] (output) {output: $\left\{\hat{\boldsymbol{\theta}}^{[f]}\right\}_{f=f_k}$};
	\path [line,transform canvas={yshift=1cm}] (localAgent) -- node {$\left\{\hat{\boldsymbol{\theta}}^{[f]}\right\}_{f=f_k}$ using (16), (17), (19) and (20)} (globalAgent) ;
	\path [line2,transform canvas={yshift=0cm}] (localAgent) -- node {$\hat{\mathbf{z}}$ using (15)} (globalAgent) ;
    \path [line,transform canvas={yshift=-1cm}] (localAgent) -- node {$\left\{\hat{\mathbf{y}}^{[f]}\right\}_{f=f_k}$ using (14)} (globalAgent) ;
    
	
	\path [doubleLine] (input) -- (localAgent);
    \path [doubleLine] (localAgent) -- (output);
  
    \path [line,transform canvas={yshift=-1cm,xshift=-3cm}] (localAgent) -- ++(0.5,0) -- node[text width=2cm,align=right] {repeat} ++(0,2) -- ++(1.1,0);
    
	\end{tikzpicture}

\caption*{ Figure 1: Operation flow and signaling exchange between the \textit{k}-th local processor and the fusion center. The three arrows in the center are performed sequentially and repeated, which corresponds to the loop from line 2 to line 10 of Table 1.
}

\end{figure}
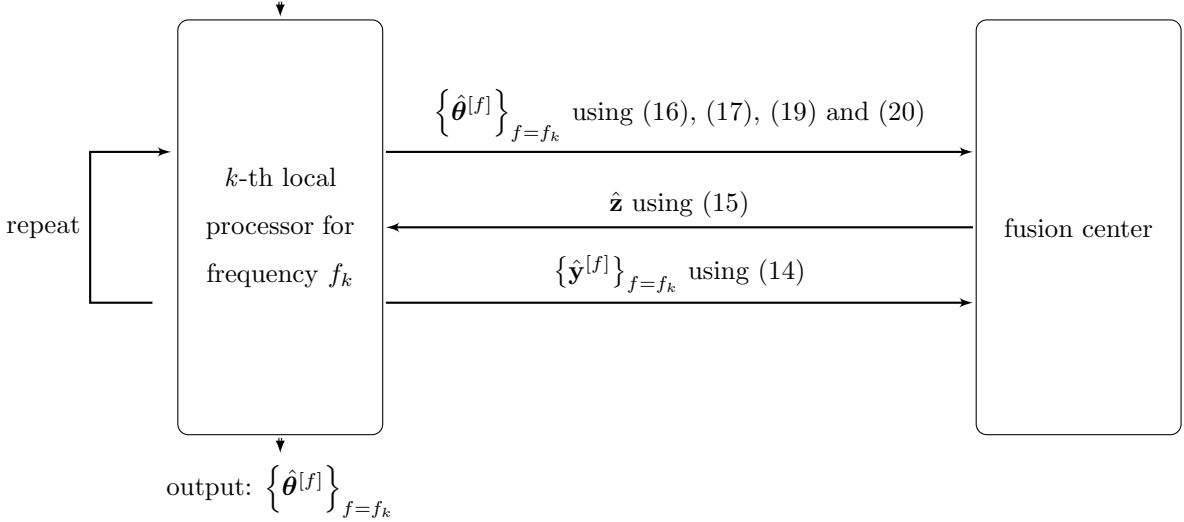

\begin{figure}[t!] 
  \centering
  \centerline{\includegraphics[width=9cm, height=7cm]{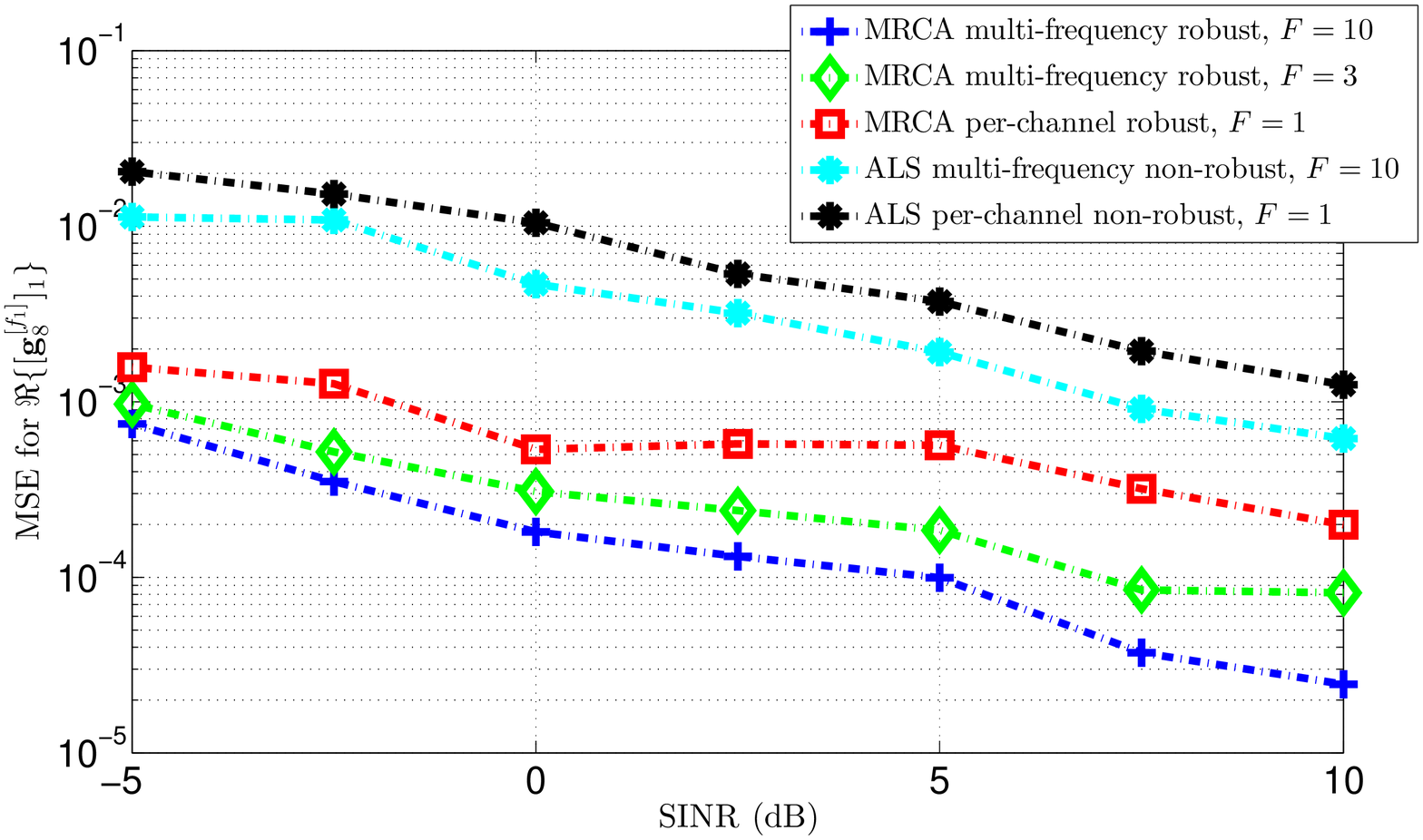}}
\caption*{ Figure 2: MSE of the real part of a given complex gain vs. SINR.} \label{fig:parameter-fix}
\end{figure}

 \begin{figure}[t!]
 \centering
\subfigure[]{\label{fig:res2-a}\includegraphics[width=7cm, height=7.5cm]{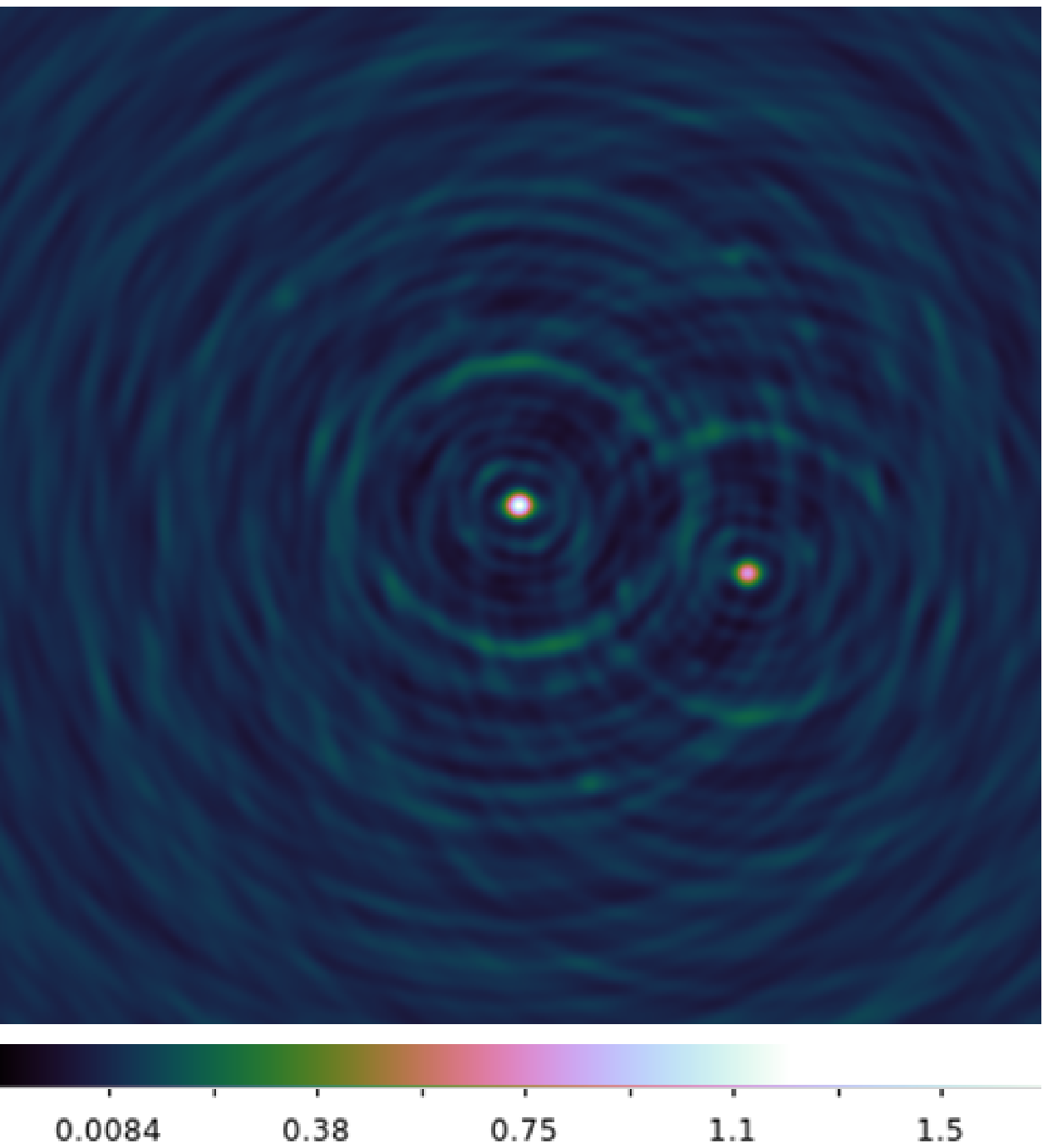}} \hspace{0.5pt} %
\subfigure[]{\label{fig:res2-b}\includegraphics[width=7cm, height=7.5cm]{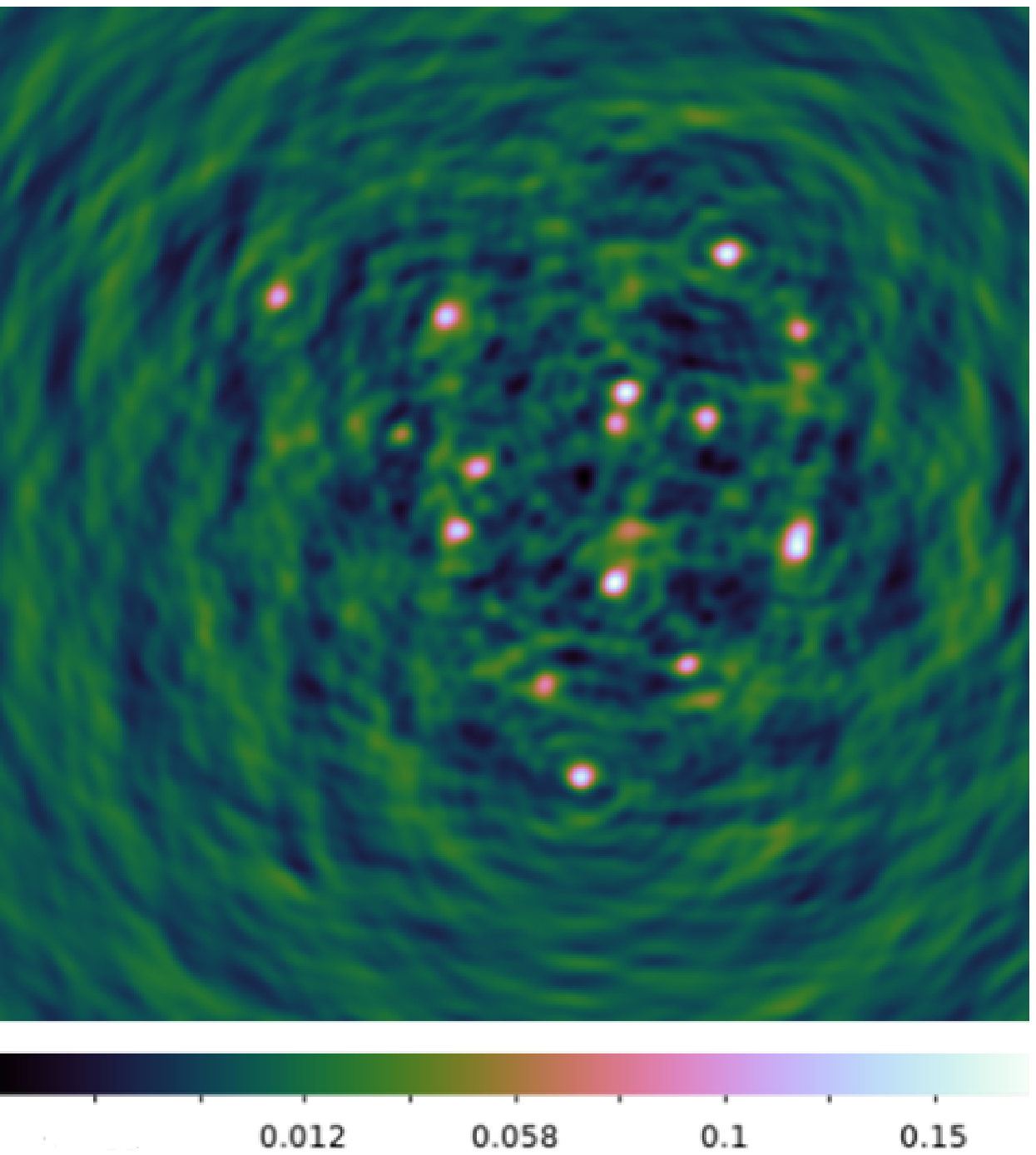}} 
\caption*{Figure 3: Dirty image representing  (a) the $D=2$ calibrator sources and $D'=16$ background sources which are hardly visible, (b) only the background sources. }
\end{figure}

\setcounter{subfigure}{0}

 \begin{figure}[t!]
 \centering
\subfigure[]{\label{fig:res3-a}\includegraphics[width=2.5cm, height=2.5cm]{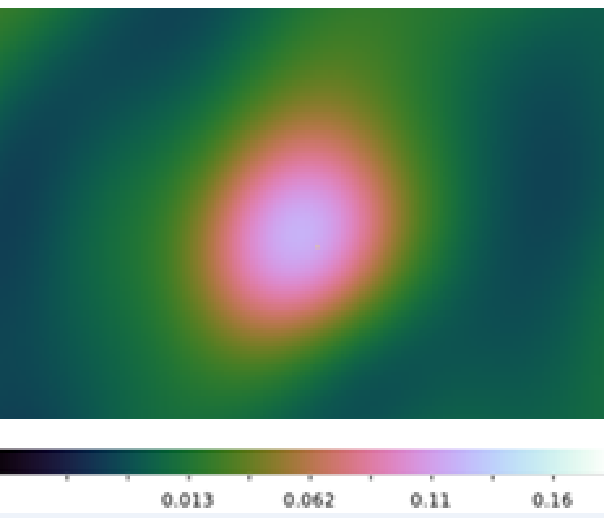}} \hspace{0.5pt} %
\subfigure[]{\label{fig:res3-b}\includegraphics[width=2.5cm, height=2.5cm]{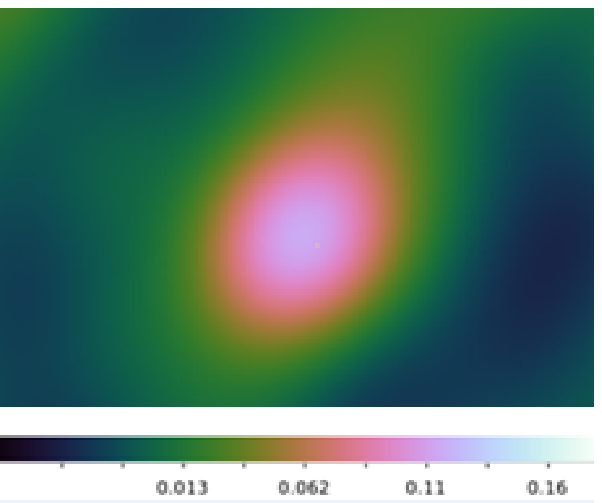}} 
\hspace{0.5pt} %
\subfigure[]{\label{fig:res3-c}\includegraphics[width=2.5cm, height=2.5cm]{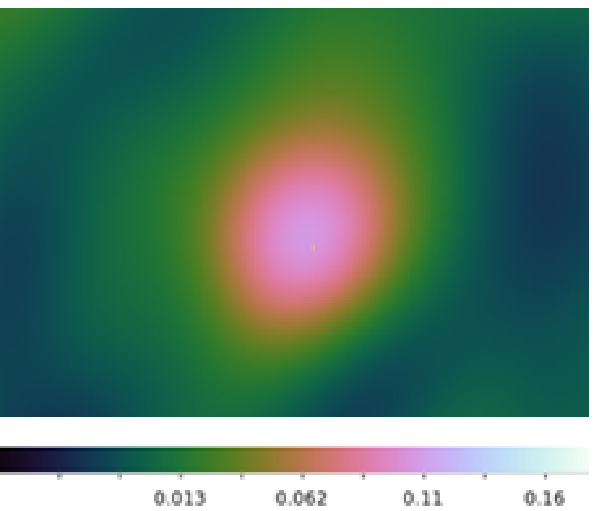}} 
\hspace{0.5pt} %
\subfigure[]{\label{fig:res3-d}\includegraphics[width=2.5cm, height=2.5cm]{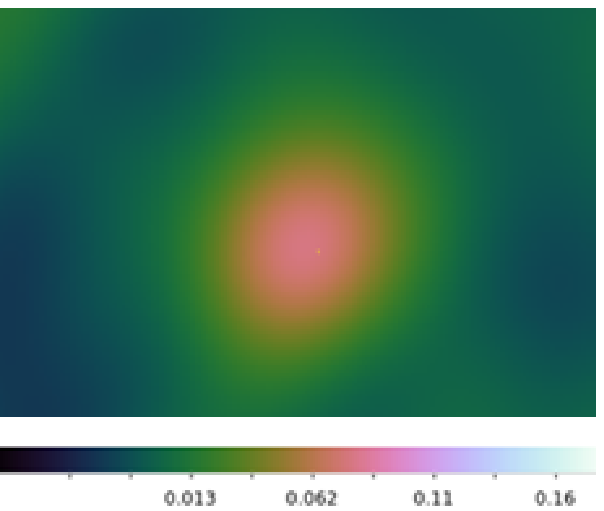}} 
\caption*{Figure 4: Corrected residual for (a) ideal case, (b) MRCA, (c) multi-frequency non-robust calibration and (d) per-channel non-robust calibration, around one of the weakest background sources. The peak flux intensity is respectively given by (a) true $0.11856$, (b) $0.117225$, (c) $0.109427$ and (d) $0.0808339$ Jy. }
\end{figure}

\setcounter{subfigure}{0}

 \begin{figure}[t!]
 \centering
\subfigure[]{\label{fig:res4-a}\includegraphics[width=2.5cm, height=2.5cm]{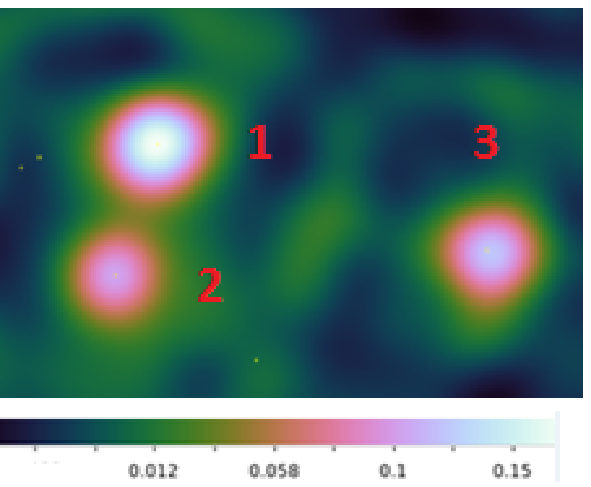}} \hspace{0.5pt} %
\subfigure[]{\label{fig:res4-b}\includegraphics[width=2.5cm, height=2.5cm]{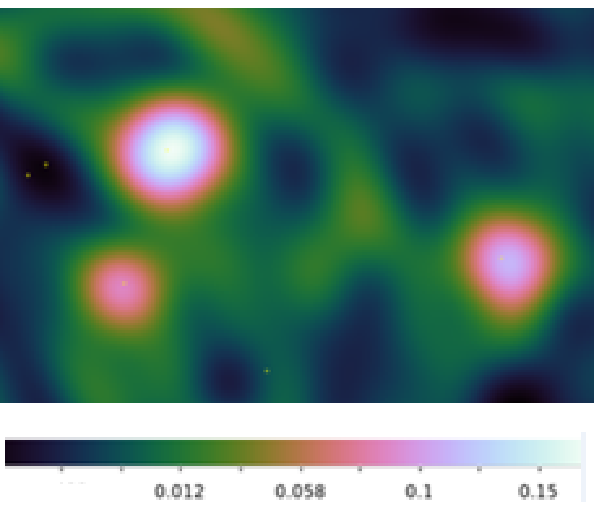}} 
\hspace{0.5pt} %
\subfigure[]{\label{fig:res4-c}\includegraphics[width=2.5cm, height=2.5cm]{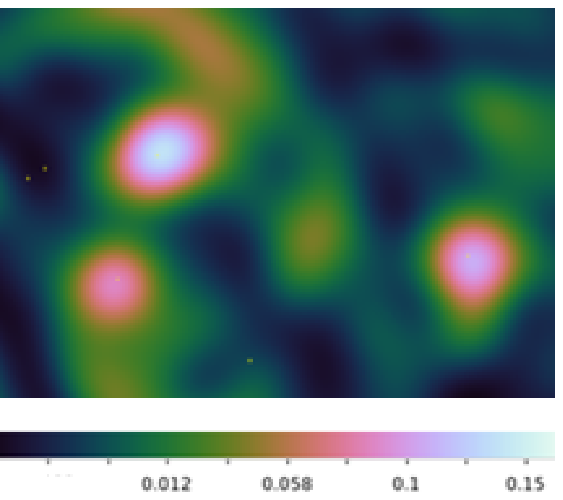}} 
\hspace{0.5pt} %
\subfigure[]{\label{fig:res4-d}\includegraphics[width=2.5cm, height=2.5cm]{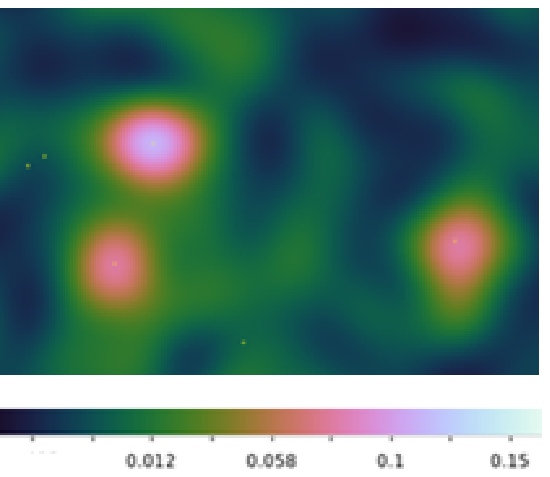}} 
\caption*{Figure 5: Corrected residual for (a) ideal case, (b) MRCA (c) multi-frequency non-robust calibration and (d) per-channel non-robust calibration in a small area surrounding three background sources. }
\end{figure}

\begin{table}[t!]
\scriptsize
\centering
\begin{threeparttable}
\caption*{\label{tab:new33}Table 2: Recovered peak flux (in Jy) for the three sources in Figure 5.}
\begin{tabular}{|l|l|l|l|} 
\hline Position & Number $1$ & Number $2$ & Number $3$  \\ \hline
 (a) True & $0.167113$ & $0.108398$ & $0.122119$ \\ \hline
  (b) & $0.162519$ & $0.0910741$  & $0.1111702$\\ \hline
  (c) & $0.136886$ & $0.0882433$  & $0.10028$\\ \hline
  (d) & $0.115676$ & $0.0831246$  & $0.082365$\\ \hline
\end{tabular}
\end{threeparttable}
\end{table}

\end{document}